\documentclass[preprintnumbers]{revtex4}
\usepackage{graphicx}
\def\ie{{\it i.e.}}
\def\eg{{\it e.g.}}

\def\etal{{\it et al.}}

\def\mpl{\ifmmode \overline M_{Pl}\else $\overline M_{Pl}$\fi}
\def\to{\rightarrow}

\begin{document}
\bibliographystyle{revtex}

\preprint{SLAC-PUB-8992/
          P3-25}

\title{Identification of Randall-Sundrum Graviton Resonances at Linear 
Colliders}

\author{Thomas G. Rizzo}

\email[]{rizzo@slac.stanford.edu}
\affiliation{Stanford Linear Accelerator Center, 
Stanford University, Stanford, California 94309 USA}

\date{\today}

\begin{abstract}
We discuss several methods which can be used to distinguish the graviton 
resonances of the Randall-Sundrum model from the graviton-like resonances 
which may occur in other theories. The Breit-Wigner line shape of the RS 
graviton is found to be particularly useful. In particular we show that the 
``effective" graviton resonance present in the model of Dvali \etal ~can be 
distinguished from those of the Randall-Sundrum scenario for a reasonably 
wide range of model parameters. 
\end{abstract}

\maketitle

\section{Introduction and Background}

Many models of new physics beyond the Standard Model 
can lead to phenomena which produce similar 
signatures at future colliders. It will thus be necessary to have tools 
available 
with which to distinguish these models and to identify the specific new 
physics source. In this report we consider the set of diagnostic tools for 
model identification of spin-2 resonances and demonstrate that the excitation 
line shape can be a useful model discriminator. 

One class of possible of new physics scenarios 
is that of extra spatial dimensions at the 
TeV scale which have been discussed in various contexts for some time{\cite
{anton}}. Amongst this class of theories is the interesting and 
phenomenologically rich model of  
Randall and Sundrum(RS){\cite {rs}} which predicts the existence of graviton 
resonances that can be produced at high energy colliders{\cite {dhr}}. Such 
resonances are easily distinguishable from other new states, such as a 
$Z'${\cite {snow}}, by measurements of the angular distribution of their decay 
products by which one can determine the spin of the initially decaying 
particle. The 
question we would like to address below is whether RS gravitons can be 
distinguished from other spin-2 states which can arise in extra dimensional 
scenarios. Here we are interested 
in the particular case where only the lightest of the RS resonances 
is accessible to detailed accelerator study, \ie, when 
the other more massive graviton excitations 
are beyond the collider center of mass energy. (Having a visible series of 
resonance states would obviously make the situation easier. For example, 
a determination that the mass spectrum of a series of spin-2 resonances 
follows the pattern of the roots of the Bessel function $J_1$ would strongly 
favor the RS interpretation.) For some scenarios the model differentiation is 
quite straightforward, \eg, in the case of Regge-like, spin-2 excitations 
of the photon and $Z${\cite{cpp}}. Here one finds that the branching fractions 
for these spin-2 states do not match those for gravitons so that these two 
models are easily separable given sufficient statistics and the visibility of 
the relevant final states at a given collider. In other cases, 
however, the situation may be more difficult, particularly so if the resonance 
appears more graviton-like. A good example of this is provided by the work of 
Carena \etal(CDLPQW){\cite {cdlpqw}} that is based on the model by 
Dvali and co-workers{\cite {dvali}} which predicts the existence of a single 
graviton-like resonance.

\section{Analysis}

\begin{figure}[htbp]
\centerline{
\includegraphics[width=5.4cm,angle=90]{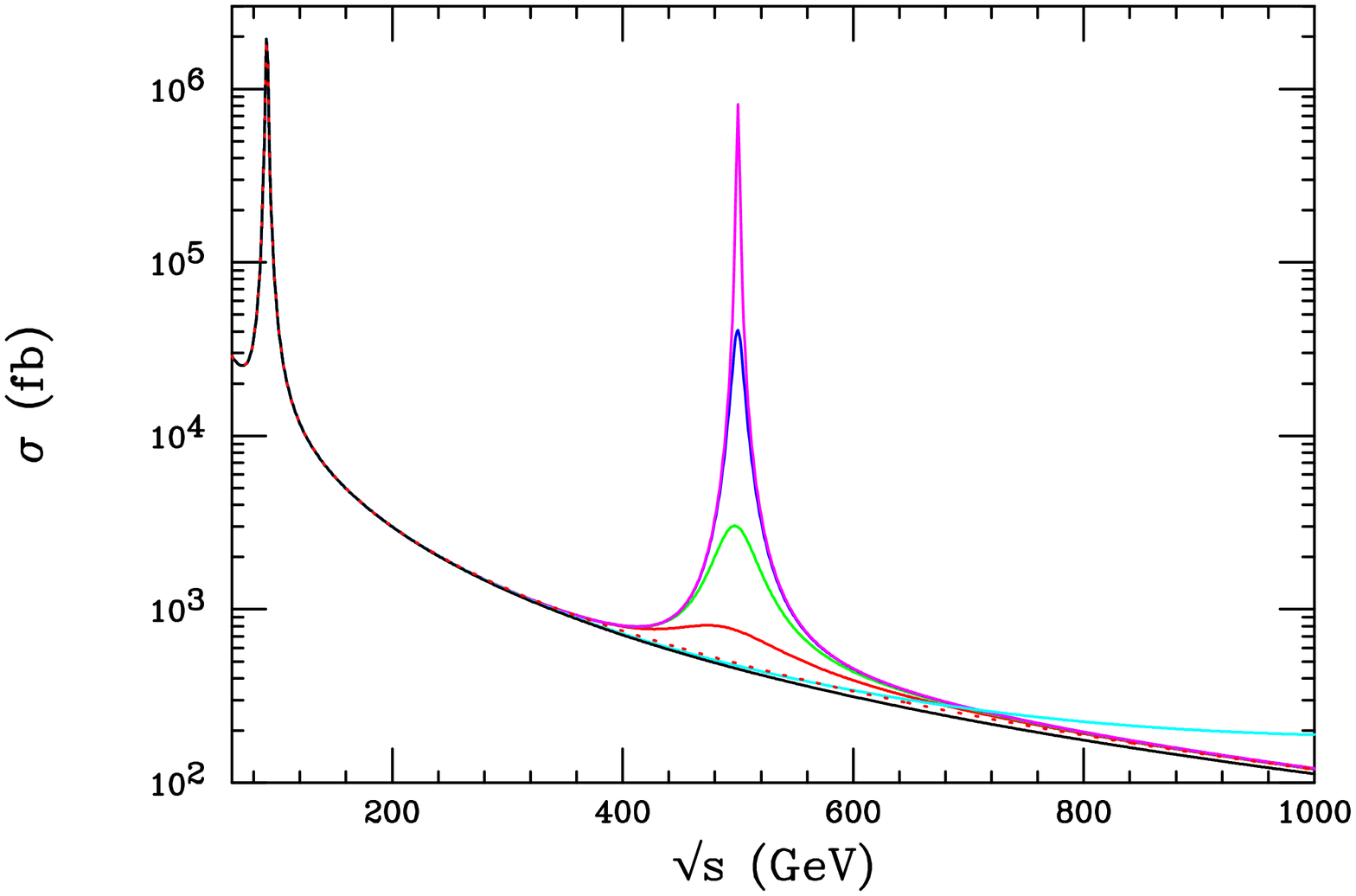}
\hspace*{5mm}
\includegraphics[width=5.2cm,angle=90]{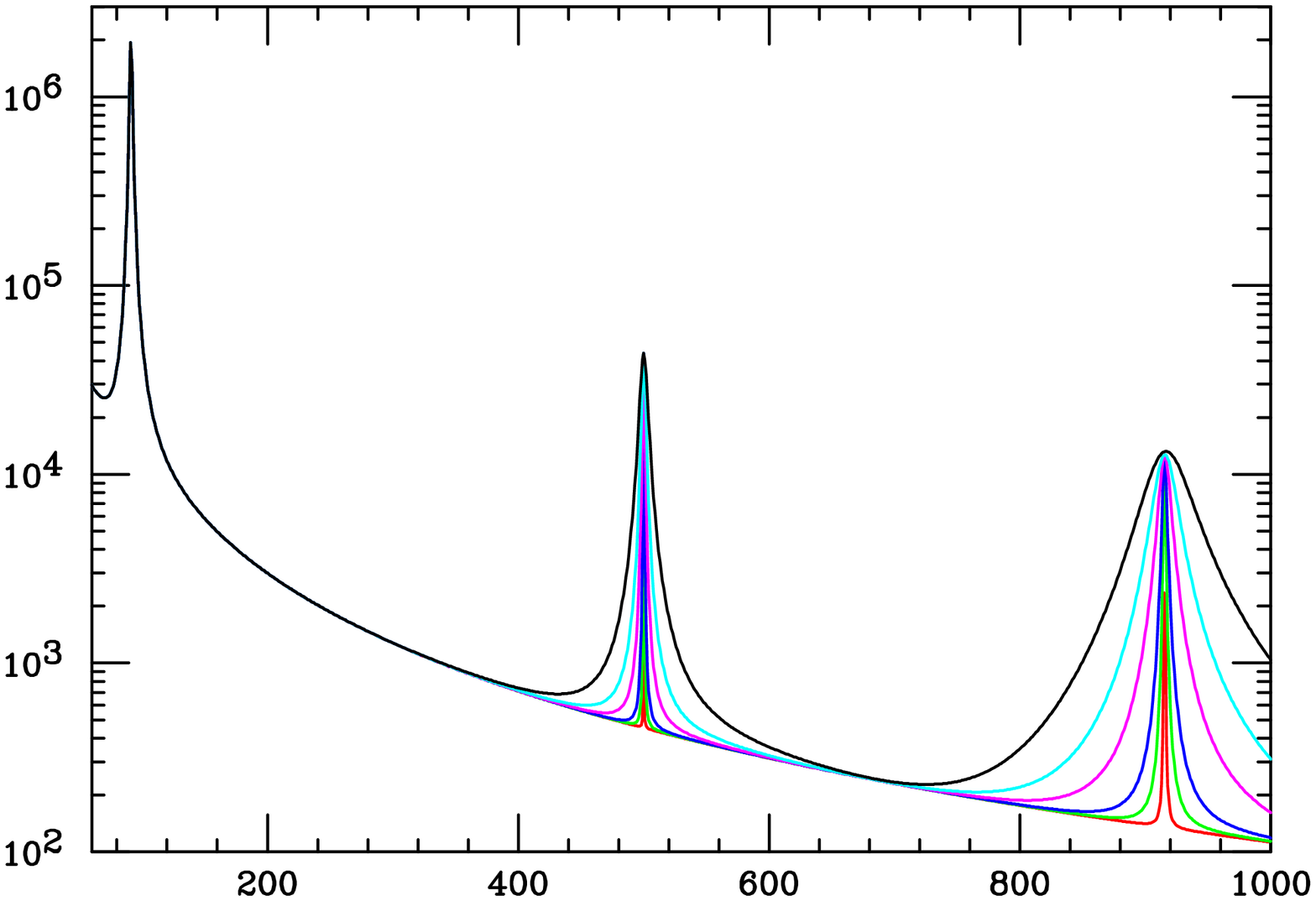}}
\vspace*{0.1cm}
\caption{$e^+e^- \to \mu^+\mu^-$ in the Dvali \etal (left) and RS(right) 
models. For the Dvali \etal ~case we assume $M_*=3$ TeV with 
$d=2$(red dots) and 
$d=3,4,5$ and 6(solid red, green, blue and magenta curves). The cyan curve is 
the ADD model prediction with constructive interference. For the RS model 
the sample curves are for the parameter $c=k/\mpl$ in the range 0.01-0.1.}
\label{p3-25_fig1}
\end{figure}

\begin{figure}[htbp]
\centerline{
\includegraphics[width=7cm,angle=90]{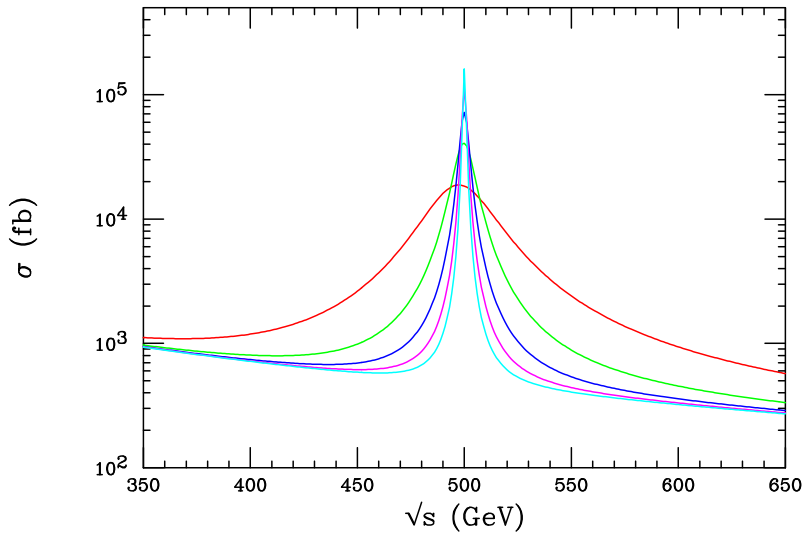}}
\vspace*{0.1cm}
\caption{Line shapes for the Dvali \etal ~graviton-like resonance in 
$e^+e^- \to \mu^+\mu^-$ assuming $d=5$, $M_d=500$ GeV and $M_*=3-6$ 
TeV in steps 
of 1 TeV corresponding to the red, green, blue, magenta and cyan curves, 
respectively.}
\label{p3-25_fig2}
\end{figure}

In this class of models the propagator of 
the graviton obtains a rather complicated structure that arises due to novel 
brane interactions, including a dimensional-dependent, as 
well as energy-dependent, imaginary part and a real part which vanishes at a 
fixed value of $s$. Hence one produces an effective resonance which is is some 
sense a ``collective" mode. Since this mode is constructed 
out of a superposition of the KK states in the graviton 
tower it has the same branching fractions as does an RS graviton and thus the 
two models cannot be distinguished using such measurements. 
The Dvali \etal ~model has 
three parameters: $d \geq 2$, the number of extra dimensions, $M_d$, the 
effective resonance mass and $M_*$, the $d-$dimensional Planck scale, which is 
on the order of a few TeV. Note that in the limit $M_d \to \infty$ we 
recover the model of Arkani-Hamed, Dimopoulos and Dvali(ADD){\cite {nima}} 
within this framework. For some values of the parameters the effective 
resonance shape looks very much like that of a relativistic Breit-Wigner(BW) 
RS graviton; this can easily be seen in 
Fig.~\ref{p3-25_fig1}. Here is shown the cross section for 
$e^+e^- \to \mu^+\mu^-$ in both the RS and Dvali \etal ~models with a common 
value chosen for the resonance masses(500 GeV) and for various values of the 
other model parameters. We note that the $d=5$ excitation curve in the 
Dvali \etal ~model is quite similar to that for the RS model with the only 
remaining free parameter 
$c=k/\mpl \simeq 0.1-0.2$. Furthermore, we note that for $d$ and $M$ 
fixed, increasing $M_*$ leads to a narrowing of the 
width and an increase in the peak height for this effective graviton 
which also makes the resonance appear more BW-like. This can be seen 
by examining the curves in Fig.~\ref{p3-25_fig2}. Can the more 
conventional shape for the RS graviton be distinguished from that of the non-BW 
``effective" resonance distribution in the Dvali \etal ~model at a Linear 
Collider?


\begin{figure}[htbp]
\centerline{
\includegraphics[width=7cm,angle=90]{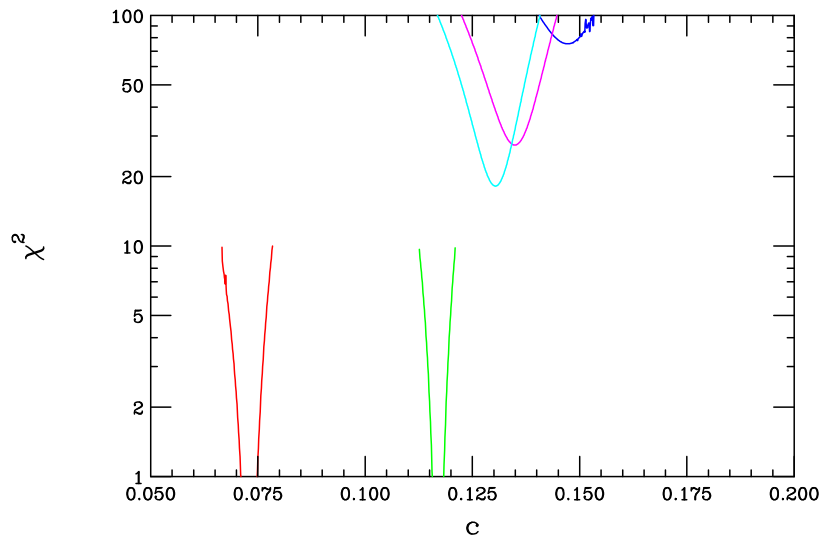}}
\vspace*{0.1cm}
\caption{Sample fits to different resonances in $e^+e^- \to \mu^+\mu^-$ 
assuming the validity of the RS 
model. The resonance mass is taken to be 1 TeV and a total integrated 
luminosity of 500$fb^{-1}$ is assumed as described in the text. The red(green) 
curves are for RS gravitons with $c=k/\mpl=0.073(0.117)$ are used as tests 
of our fitting method. The blue(magenta,cyan) curves are the corresponding 
fits for the 
Dvali \etal ~model with $M_*=5.0(5.5,5.7)$ TeV.}
\label{p3-25_fig3}
\end{figure}

To this end we undertook a preliminary study of the two line shapes which we 
imagine taking place after an unfolding of the initial state radiation and 
beamstahlung spectra; in particular we try to fit the non-BW 
Dvali \etal ~``effective" resonance under the 
assumption that it is instead a BW RS graviton and perform a fit for the RS 
model parameter $c=k/\mpl$. A poor quality fit would thus indicate that the two 
scenarios are distinguishable. In this sample study we imagine the production 
of a graviton-like resonance(\ie, spin-2 and with the proper branching 
fractions) with a mass of 1 TeV. We next generate the excitation curves for a 
set of different resonances in $e^+e^- \to \mu^+\mu^-$ assuming an integrated 
luminosity of 500 $fb^{-1}$. Since the Dvali \etal ~resonance shape is non-BW 
we do not follow the usual approach for fitting a BW but instead we fit a 
larger $\sqrt s$ region surrounding the resonance peak. 
Given the rather similar shapes of the two resonances we fit the cross section 
over the region $\sqrt s=0.80-1.32$ TeV 
in steps of 40 GeV. Since this is merely 
a first pass analysis we make no attempt to 
optimize the luminosity and assume that the 500 $fb^{-1}$ is shared equally 
amongst the generated data points; an overall 
$0.5\%$ luminosity uncertainty is 
included in this analysis. To test this approach we first try to fit 
two true RS model resonances with the input values $c=k/\mpl=0.073(0.117)$. 
This comparison is done by employing a series of `templates' which are obtained 
by generating the relevant 
cross section data for the RS model for values of $c=k/\mpl$ in the range 
0.010-0.210 in 200 steps of 0.001. We then perform our fits by using a large 
order polynomial to interpolate the values of the cross sections at other 
intermediate values of $c$ for each of the relevant values of $\sqrt s$. The 
fine granularity in the templates insures that the polynomial interpolation 
gives an extremely accurate estimate for the value of the true cross section.
These RS model fits yield $c=0.0730^{+0.0035}_{-0.0040}$ and 
$c=0.1170^{+0.0026}_{-0.0027}$, respectively, with very good $\chi^2$'s as 
shown in Fig.~\ref{p3-25_fig3}. Next, we attempt to fit the Dvali \etal ~model; 
for low values of $M_*$ the fits are very poor but improve as $M_*$ is 
increased. Fig.~\ref{p3-25_fig3} shows 
the results of these fits when $M_* \geq 5$ TeV. For cases where the value of 
$M_*$ are below 5 TeV, the resulting $\chi^2$'s are very large and are not 
shown. Note that the value of both the minimum $\chi^2$ and the fitted 
value for $c$ decrease as $M_*$ is increased. 
Clearly from this figure we see that 
for values of $M_*$ below $\simeq 5.7$ TeV the fit is 
sufficiently poor to claim that the RS hypothesis fails and the RS and Dvali 
\etal ~models are easily distinguished. In particular, for 
$M_*=5.0(5.5)$ TeV we obtain a minimum $\chi^2/d.o.f$ of 
52.9/13(27.4/13) which corresponds to a confidence level of 
$1.97\times 10^{-6}(1.70\times 10^{-2})$.  However, as we raise $M_*$ much 
beyond $\simeq 5.7$ TeV we obtain an acceptable $\chi^2$ and the two models 
are no longer distinguishable. We would expect to do somewhat better in this 
separation with an optimised distribution of luminosity; this is currently 
under investigation.

\section{Summary and Conclusions}

The above simplified analysis has shown that in some case the line shapes of 
spin-2 resonances can be a useful tool for model identification of spin-2 
resonances at linear colliders. In particular, we have shown in a toy analysis 
that the line shape can be used to distinguish the graviton resonances of the 
RS model from the ``collective" resonance present in the Dvali \etal ~model 
over a reasonable range of model parameters even without an optimization of 
the luminosity distribution in the resonance region. A detector simulation 
along the lines of the present analysis including such an optimization 
would prove useful.

%
\def\MPL #1 #2 #3 {Mod. Phys. Lett. {\bf#1},\ #2 (#3)}
\def\NPB #1 #2 #3 {Nucl. Phys. {\bf#1},\ #2 (#3)}
\def\PLB #1 #2 #3 {Phys. Lett. {\bf#1},\ #2 (#3)}
\def\PR #1 #2 #3 {Phys. Rep. {\bf#1},\ #2 (#3)}
\def\PRD #1 #2 #3 {Phys. Rev. {\bf#1},\ #2 (#3)}
\def\PRL #1 #2 #3 {Phys. Rev. Lett. {\bf#1},\ #2 (#3)}
\def\RMP #1 #2 #3 {Rev. Mod. Phys. {\bf#1},\ #2 (#3)}
\def\NIM #1 #2 #3 {Nuc. Inst. Meth. {\bf#1},\ #2 (#3)}
\def\ZPC #1 #2 #3 {Z. Phys. {\bf#1},\ #2 (#3)}
\def\EJPC #1 #2 #3 {E. Phys. J. {\bf#1},\ #2 (#3)}
\def\IJMP #1 #2 #3 {Int. J. Mod. Phys. {\bf#1},\ #2 (#3)}
\def\JHEP #1 #2 #3 {J. High En. Phys. {\bf#1},\ #2 (#3)}

\end{document}